\documentclass[aps,pre,onecolumn,notitlepage,groupedaddress]{revtex4-1}
\usepackage{epsfig,graphics,amssymb,amsmath,subeqnarray,color}
\usepackage[sfdefault=cmbr]{isomath}

\def \bnabla{\boldsymbol{\nabla}}
\def \bsigma{\boldsymbol{\sigma}}
\def \bsigmat{\tilde{\boldsymbol{\sigma}}}
\def \bu{\mathbf{u}}
\def \but{\tilde{\mathbf{u}}}

\def \bG{\mathbf{G}}
\def \be{\mathbf{e}}
\def \bT{\mathbf{T}}
\def \bx{\mathbf{x}}
\def \bf{\mathbf{f}}
\def \bn{\mathbf{n}}

\def \bx{\mathbf{x}}
\def \bxh{\hat{\mathbf{x}}}
\def \bf{\mathbf{f}}
\def \bn{\mathbf{n}}
\def \bU{\mathbf{U}}
\def \O{\mathcal{O}}
\def \deltat{\tilde{\delta}}

\begin{document}

\title{Taylor's swimming sheet: Analysis and improvement of the perturbation series}
\author{Martin Sauzade}
\author{Gwynn J. Elfring}
\author{Eric Lauga\footnote{Corresponding author. Email: \texttt{elauga@ucsd.edu}}}
\affiliation{
Department of Mechanical and Aerospace Engineering, \\
University of California San Diego, \\
9500 Gilman Drive, La Jolla CA 92093-0411, USA.}
\begin{abstract}

In G.I. Taylor's historic paper on swimming microorganisms, a two dimensional sheet was proposed  as a model for flagellated cells passing traveling waves as a means of locomotion. Using a perturbation series, Taylor computed swimming speeds up to fourth order in amplitude. Here we systematize that expansion so that it can be carried out formally to arbitrarily high order. The resultant series diverges for an order one value of the wave amplitude, but may be transformed into series with much improved  convergence properties and which yield results  comparing favorably to those obtained numerically via a boundary integral method for moderate and large values of the wave amplitudes.
\end{abstract}

\maketitle

\section{Introduction}

In his landmark 1977 paper, Purcell elucidated the unique challenges faced by microorganisms attempting to propel themselves in an inertia-less world  \cite{purcell77}. In the creeping flow limit, viscous stresses dominate, and thus shape-changing motions which are invariant under time reversal cannot produce any net locomotion -- the so-called {scallop theorem}. In order to circumvent this limitation many microorganisms are observed to pass waves along short whip-like appendages known as flagella, usually transverse planar waves for many flagellated eukaryotic cells, and helical waves for prokaryotes \cite{lighthill76,brennan77,lauga09b}.

In the first of a series of pioneering papers on the swimming of microorganisms, G.I. Taylor investigated back in 1951 such motions by considering the self propulsion of a two-dimensional sheet which passes waves of transverse displacement  \cite{taylor51}. By stipulating that such waves have a small amplitude relative to their wavelength Taylor utilized a perturbation expansion  to compute the steady swimming speed of the sheet to fourth order in amplitude. Drummond later extended Taylor's calculation of the swimming speed of an oscillating sheet to eighth order in amplitude \cite{drummond66}. A 
concise presentation of the derivation can be found in Steve Childress' textbook \cite{childress81}.

Since then many more sophisticated theoretical and computational models have been proposed to study the locomotion of microorganisms which are well documented in several review articles \cite{lighthill76,brennan77,lauga09b}. Nevertheless, the simplicity of the swimming sheet still provides opportunity for insight and analysis into such problems as swimming in viscoelastic fluids \cite{lauga07,teran10}, the synchronization of flagellated cells \cite{elfring09, elfring10}, or peristaltic pumping between walls \cite{jaffrin71,pozrikidis87b,teran08,felderhof09}. The swimming sheet has also been utilized to yield theoretical insight into inertial swimming \cite{childress81,reynolds65,tuck68}.

In this paper we show that through some mathematical manipulation the perturbation expansion for an inextensible sheet outlined by Taylor (\S\ref{series})  may be performed systematically so that the result may be obtained to arbitrary order in amplitude (\S\ref{large}). The resulting series obtained is found to be divergent for order one wave amplitudes.  Using boundary-integral computations as benchmark results (\S\ref{BI}),  we show however that the series may be transformed to obtain an infinite radius of convergence (\S\ref{comparison}), thus providing an analytical model valid for arbitrarily large wave amplitude. The coefficients for both the original and the transformed series are included as supplementary material.

\section{Series solution for Taylor's swimming sheet}
\label{series}
\subsection{Setup}

\begin{figure}
\centering
\includegraphics[width=.75\textwidth]{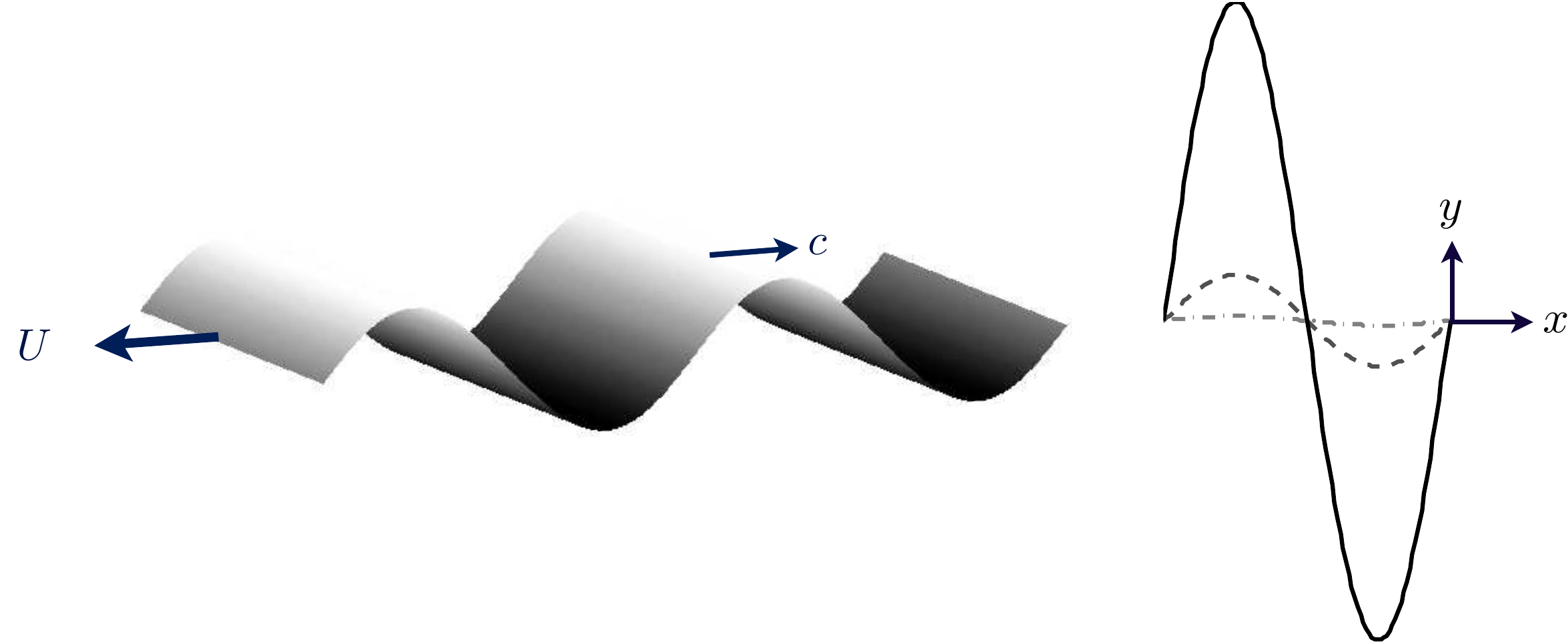}
\caption{Left: Graphical representation of the Taylor's swimming sheet. Right: Example of wave amplitude studied in this paper: $\epsilon=0.1,1,7$.}
\label{figsystem}
\end{figure}

We consider a two dimensional sheet of amplitude $b$ which passes waves of transverse displacement at speed $c=\omega/k$, where $\omega$ is the frequency and $k$ is the wavenumber (see Fig.~\ref{figsystem}). The material coordinates of such a sheet, denoted by $s$,  are given by
\begin{eqnarray}
y_{s} &=& b \sin (kx-\omega t).
\end{eqnarray}

We use the following dimensionless variables for length $x^*=xk$ and time $t^*=t\omega$ (where *'s indicate dimensionless quantities). The ratio of the amplitude of the waves to their wavelength is given by $\epsilon=bk$. For convenience we use the wave variable $z=x^*-t^*$ and therefore write 
\begin{equation}
y_{s}^* = \epsilon \sin (z) = \epsilon f (z).
\end{equation}

The regime we consider here, that of microorganisms, is the creeping flow limit governed by the Stokes equations for incompressible Newtonian flows
\begin{eqnarray}
\bnabla\cdot\bu^*&=&0, \\
\bnabla p^* &=& \nabla^2\bu^*,
\end{eqnarray}
where the velocity field $\mathbf{u^*}=\{u,v\}/c$ and pressure field $p^*=p/\mu\omega$. We now drop the *'s for convenience.

In two dimensions the continuity equation is automatically satisfied by invoking the stream function $\psi$ where
\begin{equation}
u =  - \frac{\partial \psi}{\partial y} ,\quad  v =  \frac{\partial \psi}{\partial x}\cdot
\end{equation}
The Stokes equations are then transformed into a biharmonic equation in the stream function
\begin{equation}
\nabla^{4}\psi = 0.
\end{equation}

The components of velocity of a material point of the sheet are denoted by $u_{0}$ and $v_{0}$. The conditions to be satisfied by the field $\psi$ at the surface $y = \epsilon f(z)$ are hence
\begin{equation}
-\frac{\partial \psi}{\partial y} | _{y=\epsilon f}= u_{0}, \quad   \frac{\partial \psi}{\partial x} | _{y=\epsilon f}=v_{0}.
\label{CL0}
\end{equation}

In order to find an analytical solution we seek a regular perturbation expansion in powers of $\epsilon$,
\begin{equation}
\psi \sim \sum_{k=1}^K \epsilon^{k}\psi^{(k)},
\end{equation}
with
\begin{eqnarray}
u_0\sim\sum_{k=1}^{K}\epsilon^k u_0^{(k)}, \quad v_0\sim\sum_{k=1}^{K}\epsilon^k v_0^{(k)},
\end{eqnarray}
where K is the order to which we wish to take our expansion.

We consider here only the upper-half solution which, by symmetry, is sufficient to yield the swimming velocity. The solution to the biharmonic equation which yields bounded velocities in the upper half plane, at $\O(\epsilon^k)$, is given by
\begin{equation}
\psi^{(k)} = U^{(k)}y+\sum_{j=1}^\infty \bigg[(A_{j}^{(k)}+B_{j}^{(k)}y)\sin (jz) +(C_{j}^{(k)}+D_{j}^{(k)}y)\cos (jz)\bigg]e^{-jy}.
\label{gensol}
\end{equation}
We look to solve this problem in a frame moving with the sheet and hence the terms $U^{(k)}y$ allows for the waving sheet to move relative to the far field with a velocity equal to $\bU\sim-\sum_{k=0}^{K}\epsilon^{k}U^{(k)}\be_x$.

In order to express the stream function on the boundary we expand $\psi$ in powers of $\epsilon$ about $y=0$ using Taylor expansions, and get
\begin{equation}
-\frac{\partial \psi}{\partial y}| _{y=\epsilon f} = - \sum_{k=1}^\infty \epsilon ^{k} \sum_{n=0}^{k-1}\frac{f^{n}}{n!} \frac{\partial ^{n+1} \psi ^{(k-n)}}{ \partial y^{n+1}} | _{y=0},
\end{equation}
and \begin{equation}
\frac{\partial \psi}{\partial x}| _{y=\epsilon f} = \sum_{k=1}^\infty \epsilon ^{k} \sum_{n=0}^{k-1}\frac{f^{n}}{n!} \frac{\partial ^{n+1} \psi ^{(k-n)}}{ \partial x \partial y^{n}} | _{y=0}.
\end{equation}

Substituting for $\psi$ from Eq.~\eqref{gensol} and equating with the boundary conditions we find that for $k\in [1,K]$ we must have
\begin{eqnarray}
u_0^{(k)} = -U^{(k)}- \sum_{n=0}^{k-1} \sum_{j=1}^{k-n} \frac{(-j \sin (z))^{n}}{n!}\bigg[\left(-jA_{j}^{(k-n)}+(n+1)B_{j}^{(k-n)} \right)\sin (jz) \nonumber  \\
+\left(-jC_{j}^{(k-n)}+(n+1)D_{j}^{(k-n)}\right)\cos (jz)\bigg],
\end{eqnarray}
and 
\begin{eqnarray}
v_0^{(k)} = \sum_{n=0}^{k-1} \sum_{j=1}^{k-n} \frac{(-j \sin (z)) ^{n}}{n!}&\bigg[&\left(jA_{j}^{(k-n)}-nB_{j}^{(k-n)}\right)\cos (jz) 
+\left(-jC_{j}^{(k-n)}+nD_{j}^{(k-n)}\right)\sin (jz)\bigg].
\end{eqnarray}

At order $k$, the unknowns, which are the $k^{\text{th}}$ coefficients $A_{j}^{(k)}$, $B_{j}^{(k)}$, $C_{j}^{(k)}$, $D_{j}^{(k)}$ and swimming speed $U^{(k)}$, are in the $n=0$ term only. Factoring this off and rearranging we obtain
\begin{eqnarray}\label{eq1}
u_0^{(k)}+\tilde{G}^{(k)} = -U^{(k)}+\sum_{j=1}^{k}\bigg[(jA_{j}^{(k)}-B_{j}^{(k)})\sin (jz)+(jC_{j}^{(k)}-D_{j}^{(k)})\cos (jz)\bigg],
\end{eqnarray}
and
\begin{eqnarray}\label{eq2}
v_0^{(k)}-\tilde{H}^{(k)}&=& \sum_{j=1}^{k}\bigg[jA_{j}^{(k)}\cos (jz)-jC_{j}^{(k)}\sin (jz)\bigg].
\end{eqnarray}
with 
$\tilde{G}^{(k)}$ and $\tilde{H}^{(k)}$ given by
\begin{eqnarray}
\tilde{G}^{(k)}=\sum_{n=1}^{k-1} \sum_{j=1}^{k-n} \frac{(-j \sin (z)) ^{n}}{n!}\bigg[(-jA_{j}^{(k-n)}+(n+1)B_{j}^{(k-n)} )\sin (jz)
+(-jC_{j}^{(k-n)}+(n+1)D_{j}^{(k-n)})\cos (jz) \bigg],
\end{eqnarray}
and
\begin{eqnarray}
\tilde{H}^{(k)}=\sum_{n=1}^{k-1} \sum_{j=1}^{k-n} \frac{(-j \sin (z))^{n}}{n!}&\bigg[&(jA_{j}^{(k-n)}-nB_{j}^{(k-n)} )\cos (jz) 
+(-jC_{j}^{(k-n)}+nD_{j}^{(k-n)} )\sin (jz)\bigg].
\end{eqnarray}
Provided the solution for the flow field is known for all orders up to $k-1$,  the left-hand side of Eqs.~\eqref{eq1}--\eqref{eq2} is thus known, and all the unknowns, determining the $k^{\text{th}}$ order terms, are on the right-hand side.

The terms $\tilde{G}^{(k)}$ and $\tilde{H}^{(k)}$ may conveniently  be rearranged into a Fourier series of order $k$ as
\begin{eqnarray}
\tilde{G}^{(k)}&=&\sum_{j=0}^{k}  \tilde{K}_j^{(k)} \cos (jz)+\sum_{j=1}^{k}  \tilde{S}_j^{(k)} \sin (jz),\\
\tilde{H}^{(k)}&=&\sum_{j=0}^{k}  \tilde{T}_j^{(k)} \cos (jz)+\sum_{j=1}^{k}  \tilde{R}_j^{(k)} \sin (jz).
\end{eqnarray}
A simple expression of the Fourier coefficients is not easily obtained; however, they are easily (numerically) computed.

Finally, as we show below,  the $k^{\text{th}}$ term of the components of velocity at the boundary can be written as a Fourier cosine series of order $k$
\begin{equation}
u_{0}^{(k)} =  \sum_{j=0}^{k}\alpha_{j}^{(k)}\cos(jz),
\label{u0}
\end{equation}
and 
\begin{equation}
v_{0}^{(k)} =  \sum_{j=0}^{k}\beta_{j}^{(k)}\cos(jz).
\label{v0}
\end{equation}

We can hence write, for all $k$, the system to solve
\begin{eqnarray}
jA_{j}^{(k)}-B_{j}^{(k)} &=&\tilde{S}_j^{(k)}, \\
jC_{j}^{(k)}-D_{j}^{(k)} &=&\alpha_{j}^{(k)}+\tilde{K}_j^{(k)},\\
jC_{j}^{(k)} &=&\tilde{R}_j^{(k)},\\
jA_{j}^{(k)} &=&\beta_{j}^{(k)}-\tilde{T}_j^{(k)},
\end{eqnarray}
for $j \in \left[1,k\right]$, or more compactly
\begin{eqnarray}
\mathcal{J}_j\mathbf{A}_j^{(k)}=\mathbf{\tilde{b}}_j^{(k)}.
\end{eqnarray}
The determinant of the coefficient matrix $\det(\mathcal{J}_j)=j^2$ and hence invertible $\forall j\ne 0$. The solutions for each $j$ are decoupled, and thus for each $k$ we invert a $4k$ block diagonal matrix.

Note that for the mean $j=0$ terms we obtain
\begin{eqnarray}
U^{(k)}&=& -\alpha_{0}^{(k)}-\tilde{K}_0^{(k)},\\
0&=&\beta_{0}^{(k)}-\tilde{T}_0^{(k)}.
\end{eqnarray} 
We thus see that the swimming speed at $\O(\epsilon^k)$ depends only on the mean at that order. We also find that since there is no far-field vertical velocity we require $\beta_{0}^{(k)}=\tilde{T}_0^{(k)}$, which are both known, in order to avoid an ill-posed problem. This means that since we do not allow a mean vertical flow in the solution of the stream function (which gives $\tilde{T}_0^{(k)}=0$) then the vertical boundary conditions must have zero mean, $\beta_{0}^{(k)}=0$.

Now we can solve for the swimming speed up to $O(\epsilon^k)$ by solving the above system all $k$ orders sequentially, provided we have the Fourier coefficients for the boundary conditions up to $\O(\epsilon^k)$.

\subsection{Boundary conditions}
Following Taylor \cite{taylor51}, we wish the material of the sheet to be inextensible. In a frame moving at the wave speed the shape of the sheet is at rest \cite{taylor51,childress81}, therefore in a frame moving with the sheet the boundary conditions are 
\begin{eqnarray} 
u_{0} &=& -Q \cos\theta+1, \\
v_{0} &=& -Q \sin\theta,
\end{eqnarray}
where $\tan\theta=y_s'$ and $Q$ is the material velocity in the moving frame is given by
\begin{eqnarray} 
Q &=&\frac{1}{2\pi}\int_0^{2\pi} \sqrt{1+\epsilon^2 \cos ^2 (z)} \text{d} z.
\end{eqnarray} 
Expanding in powers of $\epsilon$ and integrating we obtain
\begin{eqnarray}
Q&=&\sum_{n=0}^{\infty}\frac{(-1)^{n+1}}{(2n-1)2^{4n}}\binom{2n}{n}^{2}\epsilon^{2n},\nonumber \\
&=&\sum_{n=0}^{\infty}q_{n}\epsilon^{2n}.
\end{eqnarray}
Similarly we expand $\cos\theta$ in powers of $\epsilon$ to give
\begin{eqnarray} 
\cos\theta &=& \sum_{n=0}^{\infty}\epsilon^{2n}(-1)^{n}\frac{1}{2^{4n}}\binom{2n}{n}\left[-\binom{2n}{n}+2\sum_{r=0}^{n}\binom{2n}{n-r}\cos (2rz)\right]\nonumber \\
&=&\sum_{n=0}^{\infty}\epsilon^{2n}\sum_{r=0}^{n}t_{r}^{n}\cos(2rz).
\end{eqnarray}
Letting $k=2n$ and considering only even values we obtain 
\begin{eqnarray} 
u_{0}&=& 1-\sum_{k=0}^{\infty}\epsilon^k\sum_{r=0}^{k/2}\cos(2rz)\sum_{p=r}^{k/2}t_{r}^{p}q_{\frac{k}{2}-p}\nonumber\\
&=&-\sum_{k=2}^{\infty}\epsilon^k\sum_{r=0}^{k/2}\cos(2rz)\sum_{p=r}^{k/2}t_{r}^{p}q_{\frac{k}{2}-p}.
\end{eqnarray}
Now letting $j=2r$ we find for even $j$ and even $k\ge2$
\begin{equation}
u_0^{(k)}=\sum_{j=0}^{k}\alpha_j^{(k)}\cos(jz),
\end{equation}
where we have defined
\begin{equation}
\alpha_j^{(k)}=-\sum_{p=j/2}^{k/2}t_{\frac{j}{2}}^p q_{\frac{k}{2}-p},
\end{equation}
while $u_0^{(k)}=0$ for odd $k$ and $\alpha_j^{(k)}=0$ for odd $j$.

We then know that
\begin{eqnarray}
v_0=-y_s'(z)Q\cos\theta,
\end{eqnarray}
and hence we find for odd $j$ and odd $k\ge3$
\begin{eqnarray}
v_j^{(k)}&=& \sum_{j=1}^{k}\beta_j^{(k)}\cos(jz),\\
\beta_1^{(k)}&=& \alpha_0^{(k-1)}+\frac{1}{2}\alpha_2^{(k-1)},\\
\beta_j^{(k)}&=& \frac{\alpha_{j-1}^{(k-1)}+\alpha_{j+1}^{(k-1)}}{2}, \quad 3\le j\le k-2,\\
\beta_k^{(k)}&=&\frac{1}{2}\alpha_{k-1}^{(k-1)},
\end{eqnarray}
and for $k=1$ $\beta_1^{(1)}=-1$. In contrast, $v_0^{(k)}=0$ for even $k$ and $\alpha_j^{(k)}=0$ for even $j$. We see that the vertical component of the boundary velocity has no mean component at any order in $\epsilon$ which, as we saw in the previous section, is required given the form of the solution. With these coefficients we can now solve a linear system at each order to obtain $U^{(k)}$ to arbitrary order.

In practice the number of terms obtainable will be limited by numerical technique. To obtain the first one thousand terms of the series used in the analysis in the following sections, the system of equations was solved using the $\textit{C}$ programming language with GNU MP, the {GNU multiple precision arithmetic library} \cite{gmp}, using $300$ digits of accuracy.

\section{Analysis and improvement of the perturbation series}
\label{large}
In the previous sections we presented methodology to obtain the solution to the swimming speed $U$ in the form of a perturbation series
\begin{equation}\label{originalseries}
U(\epsilon)\sim\sum_{k=1}^{K}U^{(k)}\epsilon^{k}.
\end{equation}
It remains of course to be seen whether the series will converge to $U$ for arbitrary $\epsilon$. 
We analyze here the convergence properties of the series, and methods to improve upon that convergence.

\begin{figure}
\centering
\includegraphics[width=0.75\textwidth]{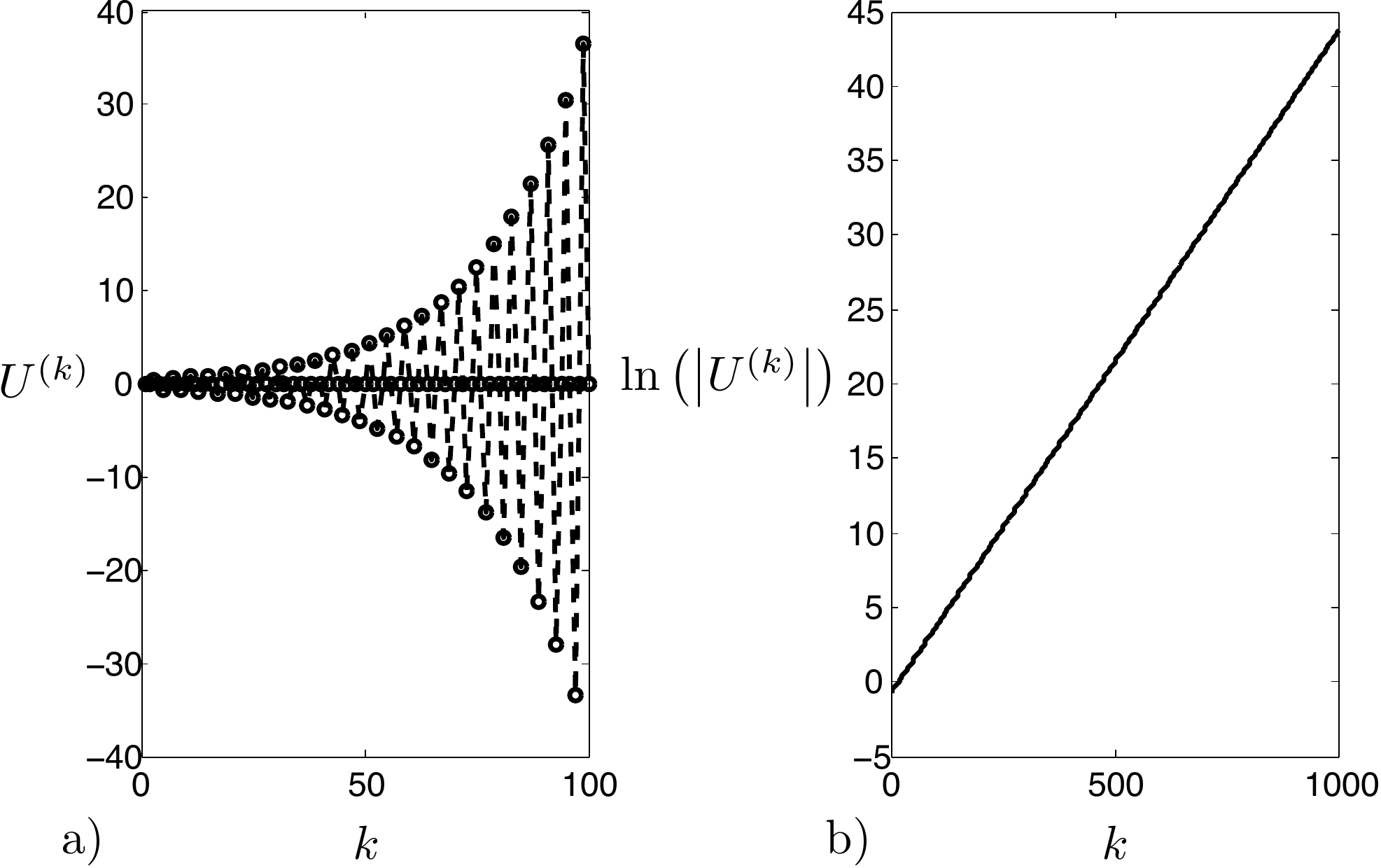}
\caption{Coefficients of the series for the swimming speed, Eq.~\eqref{originalseries}.  
(a): The first 100 terms of the series $U^{(k)}$; (b): $\ln\left(
\left|U^{(k)}\right|\right)$ for $k=1$ to $1000$ for nonzero values of  $U^{(k)}$.}
\label{fig2_3label}
\end{figure}

In Fig.~\ref{fig2_3label} we plot the coefficients of the series $U^{(k)}$ against $k$. On Fig.~\ref{fig2_3label}a  are plotted the first 100 terms, and on Fig.~\ref{fig2_3label}b the logarithm of the absolute value of the nonzero terms up to $k=1000$.
We see that the coefficients have an exponentially increasing amplitude while alternating in sign, $U^{(k)}>0$ for $k=4n-2$ and $U^{(k)}<0$ for $k=4n$ where $n\in\mathbb{N}$. We also note that due to the $\epsilon\rightarrow-\epsilon$ symmetry of the geometry in the problem, all odd powers in the series are zero. It is therefore useful to recast the series as follows
\begin{eqnarray}\label{newseries}
U=
\sum_{k=1}^{K}U^{(k)}\epsilon^{k}=\sum_{k=1}^{K/2}U^{(2k)}\epsilon^{2k}= \delta\sum_{k=0}^{K/2-1}c_k\delta^k,
\end{eqnarray}
where $c_k=U^{(2k+2)}$ and $\delta=\epsilon^2$. The coefficients $c_k$ for $k=1$ to 500 have been reproduced in the included supplementary material of the manuscript.

The sign of $c_k$ alternates in a regular manner which indicates that the nearest singularity lies on the negative real axis and since only positive values of $\delta$ have any meaning, there is no physical significance to the singularity;  it does of course govern the radius of convergence of the series \cite{vandyke74}.

\subsection{Series convergence}
\begin{figure}
\centering
\includegraphics[width=0.7\textwidth]{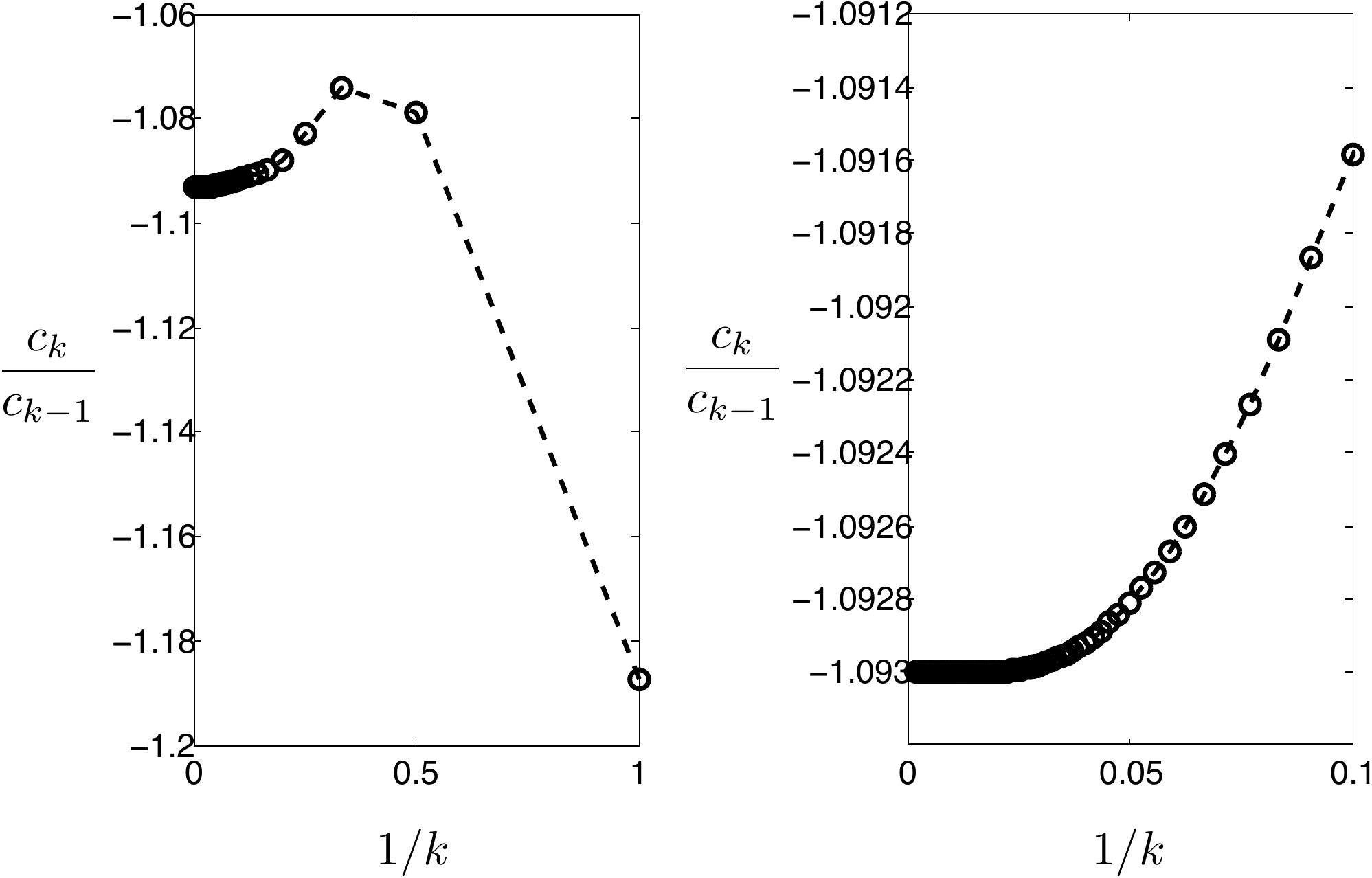}
\caption{Domb-Sykes plot of the coefficients $c_k$, from the series in  Eq.~\eqref{newseries}, shows convergence to $1/\delta_0\approx -1.093$.}
\label{figckDS}
\end{figure}

The radius of convergence, $\delta_0$, of the power series
\begin{eqnarray}
f(\delta)\sim\sum_k c_k \delta^k,
\end{eqnarray}
may be simply found by using the ratio test
\begin{eqnarray}
\delta_0 = \lim_{k\rightarrow \infty} \frac{c_{k-1}}{c_k}\cdot
\end{eqnarray}
In order to find this value we must extrapolate due to the finite number of terms. In order to aid this process Domb and Sykes noted it is helpful to plot $c_k/c_{k-1}$ against $1/k$ \cite{domb57}. The reason is that if the singular function $f$, has a dominant factor
\begin{eqnarray}
(\delta_0-\delta)^\gamma \quad &\text{for}& \quad \gamma\ne 0,1,2,...,\\
(\delta_0-\delta)^\gamma\ln(\delta_0-\delta) \quad &\text{for}& \quad \gamma= 0,1,2,...,
\end{eqnarray}
then the coefficients behave like
\begin{equation}
\frac{c_{k}}{c_{k-1}}\sim \frac{1}{\delta_{0}}\left(1-\frac{1+\gamma}{k}\right),
\label{dombsykes}
\end{equation}
for large $k$ \cite{vandyke74,hinch91}. The result in Eq.~(\ref{dombsykes}) indicates that the intercept $1/k=0$ in a Domb-Sykes plot gives the reciprocal of the radius of convergence while the slope approaching the intercept gives $\gamma$. In Fig.~\ref{figckDS} we show the Domb-Sykes plot of the series $c_k$. The plot indicates that the nearest singularity is at $\delta_{0}\approx-0.914912217581184$, and that $\gamma=-1$ corresponding to  a first order pole. 

\subsection{Euler transformation}\label{sec:euler}

One approach to improve convergence of the series is to factor out the first-order pole characterized above, and characterize the singularities of the new series.  However we find that that series is no more tractable due to the presence of an apparent branch cut in the complex plane close to $\delta = -1$.

\begin{figure}
\centering
\includegraphics[width=0.7\textwidth]{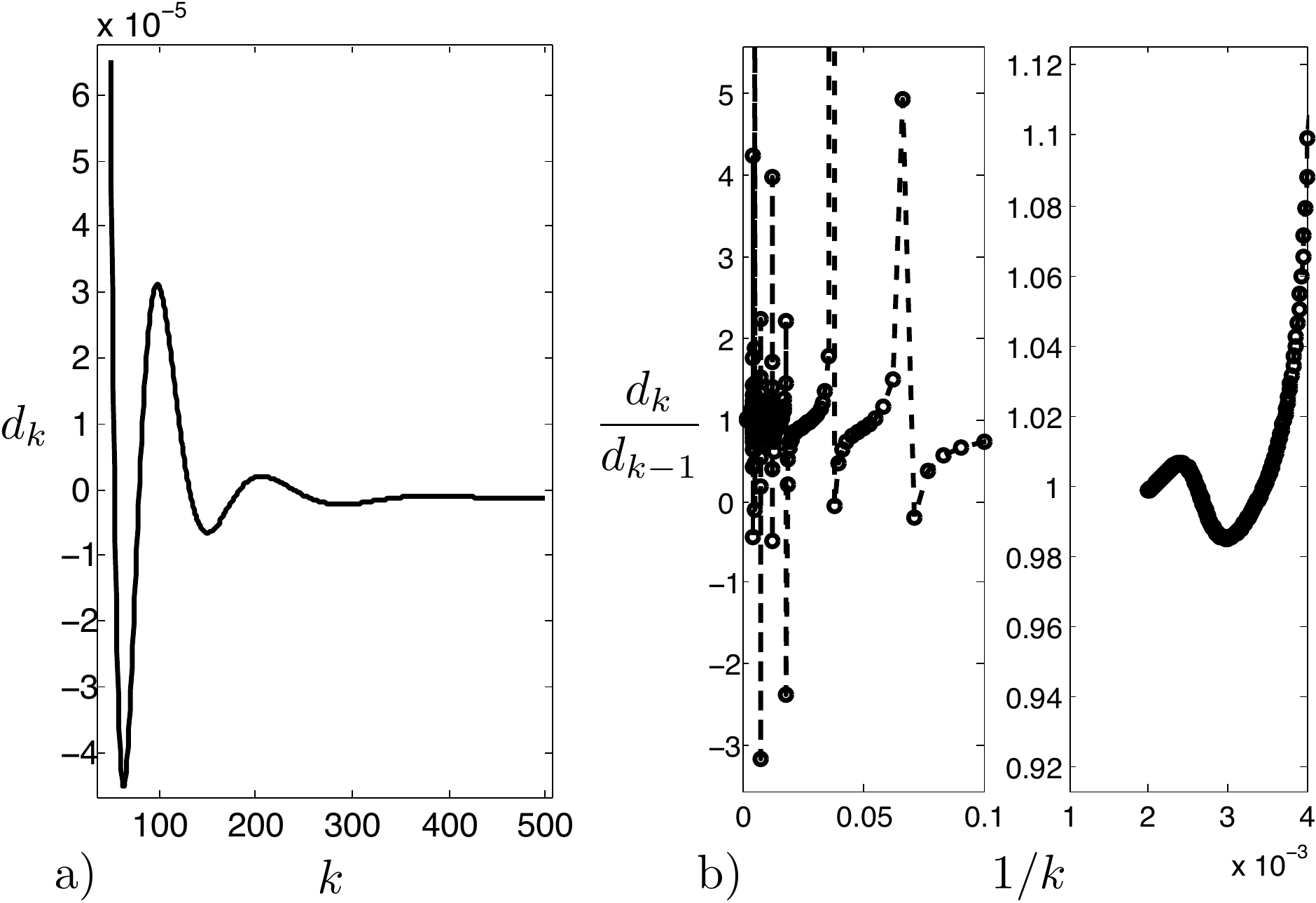}
\caption{
Coefficients of the new series for the  swimming speed using  the Euler transformation, Eq.~\eqref{eq:euler}.  
(a) Coefficients $d_{k}$ of the new series; 
(b) Domb-Sykes plot of the coefficients shows a convergence to one.}
\label{figdk}
\end{figure}

Alternatively, the original non-physical singularity $\delta_0$ may be mapped to infinity using a Euler transformation and introducing a new small variable
\begin{eqnarray}
\deltat=\frac{\delta}{\delta-\delta_0}\cdot
\end{eqnarray}
The power series for $f$ is then recast as
\begin{equation}\label{eq:euler}
f\sim\sum_k c_k \delta^k \sim \sum_k d_k \deltat^k.
\end{equation}
The coefficients $d_k$ for $k=1$ to 500 have been reproduced in the included supplementary material of the manuscript. Their values for $k>50$ are shown in Fig.~\ref{figdk}a and we can see that they decay in magnitude for large $k$. 

In order to find the radius of convergence of the new series, $d_k$, we again turn to the Domb-Sykes plot, which is show in Fig.~\ref{figdk}b. We see that it appears $d_k/d_{k-1}\rightarrow 1$ as $k^{-1}\rightarrow 0$ and since $\delta/(\delta-\delta_0)\rightarrow 1$ when $\delta\rightarrow \infty$, we have now have an infinite radius of convergence in the original variable $\delta$. Note that the vastly improved convergence does not necessarily mean the series will provide a good approximation beyond $\delta_0$ \cite{hinch91}; however we will see in the results section that it actually provides an excellent fit to the numerical results.

\subsection{Pad\'{e} approximants}\label{sec:pade}

\begin{figure}[t]
\centering
\includegraphics[width=0.75\textwidth]{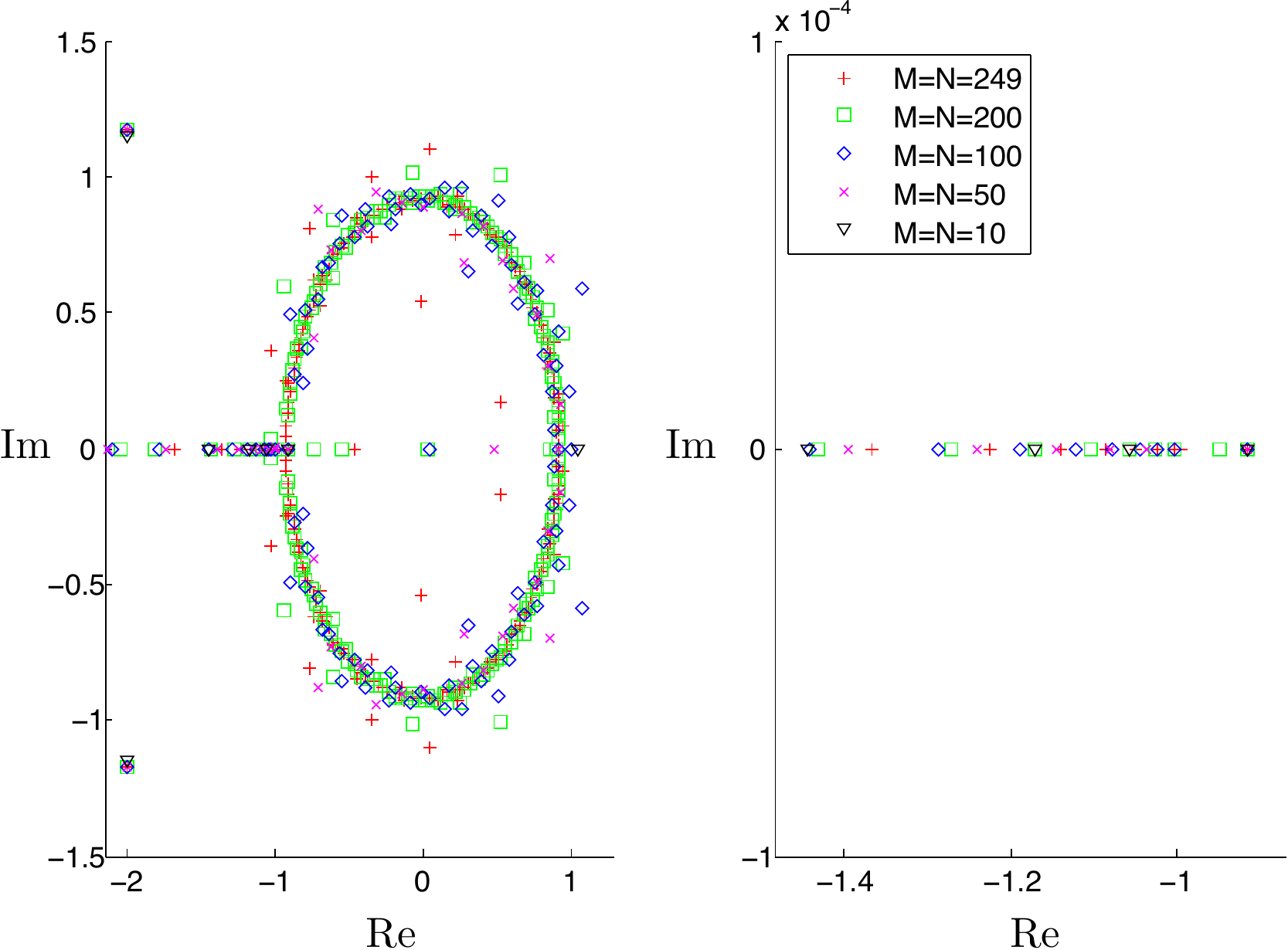}
\caption{Zeros in the complex plane of the denominators of various  Pad\'{e} approximants for $M=N=10$, 50, 100, 200 and 249.}
\label{figpadezero}
\end{figure}

A popular scheme to improve the convergence properties of series is to recast the series as a rational polynomial
\begin{eqnarray}
f(\delta)\sim \sum_{k=0}^{M+N} c_k\delta^k \sim \frac{\sum_0^M a_k \delta^k}{\sum_0^N b_k\delta^k}=P_{N}^{M},
\end{eqnarray}
where $M+N\le K/2-1 $. If we multiply both sides by the denominator $\sum b_k\delta^k$ for the terms of order $\delta^k$ where $k=M+1:M+N$ we obtain a square matrix to invert for $b_1,...,b_N$ and one takes $b_0=1$ with no loss of generality \cite{bender78}. One can then solve for $a_k$.

We apply this method to our swimming sheet, and plot the zeros of different Pad\'{e} denominators with $M=N$ in Fig.~\ref{figpadezero}. 
It is evident that the pole we identified earlier at  $\delta_{0}$ is well reproduced here. The interesting feature beyond this is the fact that the remaining zeros do not exhibit any consistency, which indicates a branch cut in the complex plane.

\subsection{Shanks transformation}\label{sec:shanks}

A scheme to improve the rate of convergence of a sequence of partial sums
\begin{eqnarray}
S_n=\sum_{k=0}^{n}c_k\delta^k,
\end{eqnarray}
for $n=0$ to $N\le K/2-1$, is to assume they are in a geometric progression
\begin{eqnarray}
S_n=A+BC^n.
\end{eqnarray}
Solving for A by nonlinear extrapolation of three sums yields
\begin{eqnarray}
A_n=S_n-\frac{\left(S_{n+1}-S_{n}\right)\left(S_n-S_{n-1}\right)}{\left(S_{n+1}-S_n\right)-\left(S_n-S_{n-1}\right)}.
\end{eqnarray}
The $A_n$'s for $n=1$ to $N-1$, can then be considered a series of partial sums and the Shanks transformation may be thereby repeated $(N-1)/2$ times \cite{hinch91}.

\section{Boundary Integral Formulation}
\label{BI}
In order to provide benchmark results for the analysis of the perturbation series and its various transformations, we use the boundary integral method to obtain what we will consider to be an {exact} solution of the swimming speed for waves of arbitrarily large amplitude.

We briefly summarize the principle of the method here. The Lorentz reciprocal theorem states that two solutions to the Stokes equations, $\left(\bu, \bsigma\right)$ and $\left(\but,\bsigmat\right)$ are related by
\begin{eqnarray}
\int_S \left(\bu\cdot\bsigmat\right)\cdot\bn \ dS=\int_S \left(\but\cdot\bsigma\right)\cdot\bn \ dS,
\label{lorenz}
\end{eqnarray}
within a volume $V$ bounded by the surface $S$ whose unit normal $\bn$ is taken pointing into the fluid. The velocity and stress fields, $\but(\bx)$ and $\bsigmat(\bx)$, are taken to be fundamental solutions for two-dimensional Stokes flow due to a point force at $\bx_0$,
\begin{eqnarray}
\but(\bx)&=&\frac{1}{4\pi}\bG(\hat{\bx})\cdot\tilde{\bf}(\bx_0), \\
\bsigmat(\bx)&=&\frac{1}{4\pi}\bT(\hat{\bx})\cdot\tilde{\bf}(\bx_0),
\end{eqnarray}
where $\hat{\bx}=\bx-\bx_0$ and the two dimensional Stokeslet $\bG$, and stresslet $\bT$ are given by
\begin{eqnarray}
\bG&=&-\mathbf{I}\ln(\left|\bxh\right|)+\frac{\bxh\bxh}{\left|\bxh\right|^2},\\
\bT&=&-4\frac{\hat{\bx}\hat{\bx}\hat{\bx}}{\left|\hat{\bx}\right|^4}\cdot
\end{eqnarray}
Taking the singular point $\bx_0$ to be on the boundary $S$ one obtains from Eq.~\eqref{lorenz} a boundary integral solution to two-dimensional Stokes equations for the velocity
\begin{eqnarray}
\bu(\bx_0)=\frac{1}{2\pi}\int_S\left(\bu(\bx)\cdot\bT(\hat{\bx})\cdot\bn(\bx)-\bf(\bx)\cdot\bG(\hat{\bx})\right)\ d S(\bx), \label{bi2}
\end{eqnarray}
where $\bf=\bsigma\cdot\bn$.

We wish to capture the swimming speed of an infinite sheet therefore the domain of integration is an entire half plane of fluid bounded by the sheet. In order to avoid performing an integration over the entire bound it is convenient to use an array of periodically placed Stokeslets and stresslets, given by
\begin{eqnarray}
\bG^p&=&\sum_{n=-\infty}^{\infty}-\mathbf{I}\ln(\left|\hat{\bx}_n\right|)+\frac{\hat{\bx}_n\hat{\bx}_n}{\left|\hat{\bx}_n\right|^2},\\
\bT^p&=&\sum_{n=-\infty}^{\infty}-4\frac{\hat{\bx}_n\hat{\bx}_n\hat{\bx}_n}{\left|\hat{\bx}_n\right|^4},
\end{eqnarray}
where $\hat{\bx}_n=\{\hat{x}_0+2\pi n,\hat{y}_0\}$, so that we may then instead integrate $\bG^p$ and $\bT^p$ over a single period \cite{pozrikidis87}. The periodic Stokeslet and stresslet may be conveniently expressed in closed form \cite{pozrikidis87,pozrikidis92}, through the use of the following summation formula
\begin{eqnarray}
A=\sum_{n=-\infty}^{\infty}\ln(\left|\hat{\bx}_n\right|)=\frac{1}{2}\ln\left[2\cosh(\hat{y}_0)-2\cos(\hat{x}_0)\right],
\end{eqnarray}
and its derivatives, as follows
\begin{eqnarray}
G_{xx}^p &=& -A-\partial_yA+1, \\
G_{xy}^p &=& y\partial_xA, \\
G_{yy}^p &=& -A+y\partial_y A,
\end{eqnarray}
and
\begin{eqnarray}
T^p_{xxx} &=& -2\partial_x(2A+y\partial_yA), \\
T^p_{xxy} &=& -2\partial_y(y\partial_yA), \\
T^p_{xyy} &=& 2y\partial_{xy}A, \\
T^p_{yyy} &=& -2(\partial_yA-y\partial_{yy}A).
\end{eqnarray}
The remaining elements follow from a permutation of the indices of the Stokeslet and stresslet which leaves the right hand side unchanged \cite{pozrikidis92}.

The flow is quiescent at infinity and periodic on $2\pi$ and therefore the domain of integration $S$ reduces to the surface of the sheet over one period. To facilitate integration the continuous boundary is discretized into $N$ straight line elements $S_n$ and we assume that $\bf$ is a linear function over each particular interval, $\bf\rightarrow\bf_n$ (see Ref.~\cite{higdon85}). We decompose the boundary velocity into surface deformations and rigid body motion $\bu\rightarrow\bu_n+\bU$, where $\bu_n$ is a linear function over each interval and $\bU\equiv -U\be_x$. Then $\bx_0$ is taken at the center of each of the the $N$ segments $S_n$,where the velocity is known, $\bx_0\rightarrow \bx_m$. The $\bG^p$ and $\bT^p$ are regularized by subtracting off the Stokeslet and stresslet from their periodic counterparts. The two-dimensional Stokeslet and stresslet are then integrated analytically and added back. 

We thereby obtain from Eq.~\eqref{bi2} a linear system for $\bf_n$ and $U$, given by
\begin{eqnarray}
\bu(\bx_m)+\bU=\frac{1}{2\pi}\sum_{n=1}^N\Bigg[-\int_{S_n}\bf_n\cdot\left(\bG^p-\bG\right)d S_n 
-\int_{S_n}\bf_n\cdot\bG dS_n
\nonumber\\
+\int_{S_n}(\bu_n+\bU)\cdot\left(\bT^p-\bT\right)\cdot\bn_n d S_n
+\int_{S_n}(\bu_n+\bU)\cdot\bT\cdot\bn_n dS_n
\Bigg].
\end{eqnarray}

We then obtain $U$ by specifying that the sheet is force free
\begin{eqnarray}
\sum_{n=1}^N \left[\be_x\cdot \int_{S_n} \bf_n dS_n\right]=0.
\end{eqnarray}

The numerical procedure was validated by reproducing Pozrikidis' results for shear flow over sinusoidal surface \cite{pozrikidis87}.

\section{Comparison between series solution and computations}
\label{comparison}

\subsection{Series solution}

\begin{figure}
\centering
\includegraphics[width=0.5\textwidth]{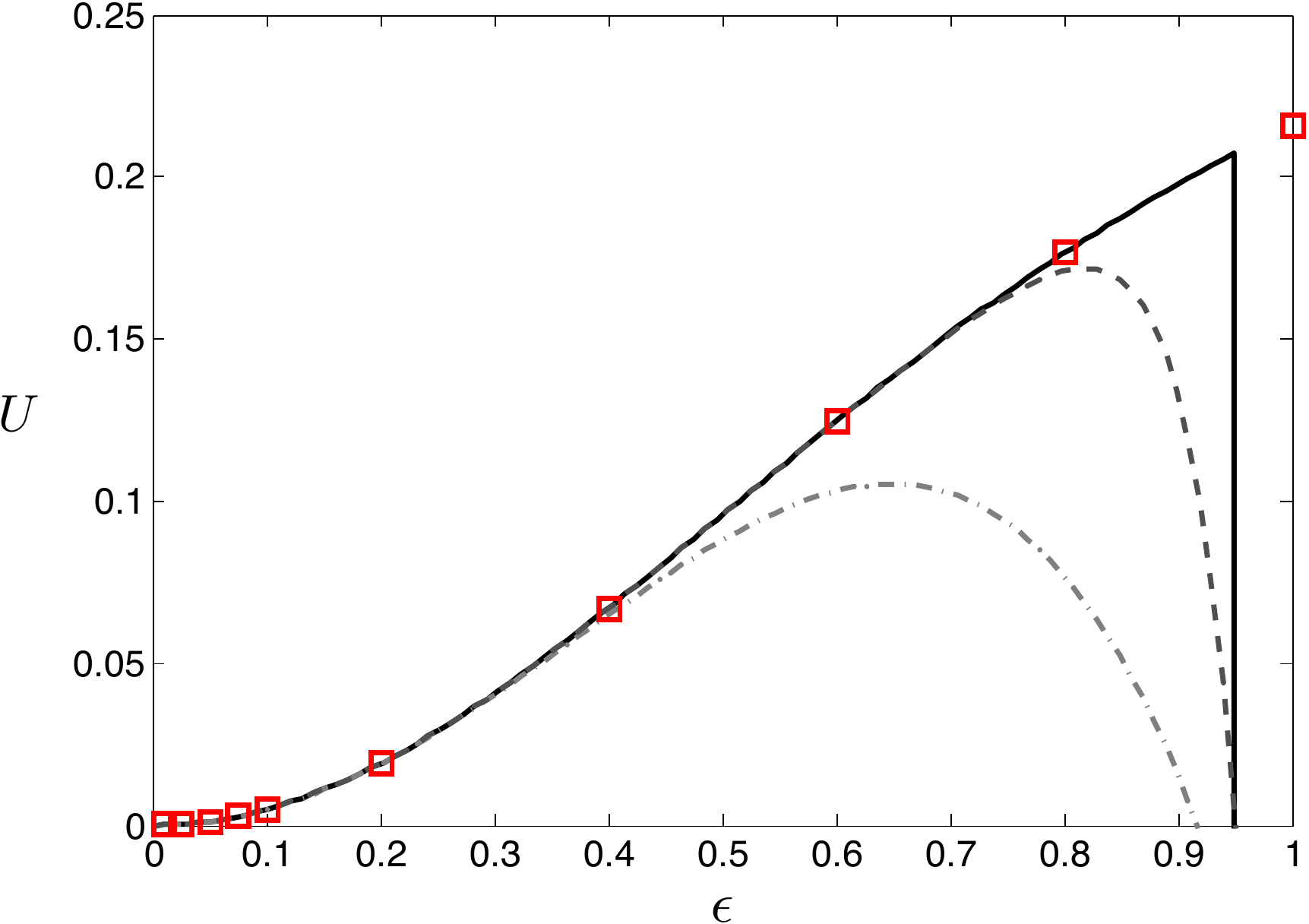}
\caption{Swimming speed, $U$, against wave amplitude, $\epsilon$, for the unaltered series, Eq.~\eqref{originalseries},  with $K=4$ (dashed-dot), $K=20$ (dashed), $K=1000$ (solid). The series diverges for $\epsilon\approx 0.9565$. Red squares indicate data points from the boundary integral method.}
\label{figseries}
\end{figure}

We first show the convergence of the unaltered series expansion, Eq.~\eqref{originalseries}, in Fig.~\ref{figseries} where we display the swimming speed of the sheet, $U$, as a function of its amplitude, $\epsilon$.
The red squares indicate numerical points computed with the boundary integral method. We
 plot the results for Taylor's original fourth order expansion (dashed-dot line) which is reasonably accurate up to $\epsilon\approx 0.4$. The series with  $K=20$ is shown in dashed line. As we add terms, we get that the series with $K=1000$ (solid line)  fails to converge beyond the singularity at $\epsilon=\sqrt{-\delta_{0}}\approx0.95651$, as expected from the analysis in \S\ref{large}. 

\subsection{Euler transformation}
\begin{figure}
\centering
\includegraphics[width=0.5\textwidth]{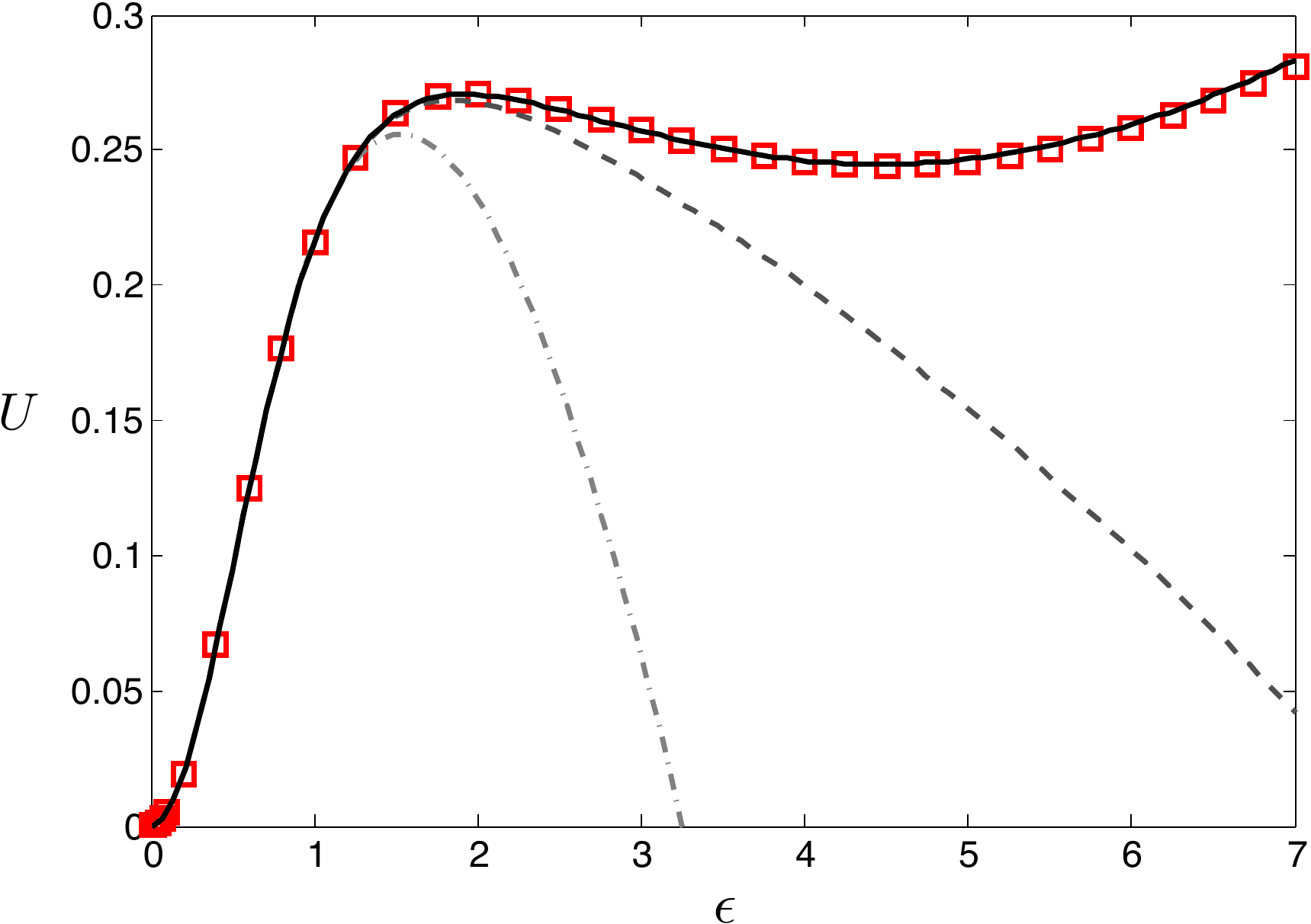}
\caption{Swimming speed, $U$, against wave amplitude, $\epsilon$, for the Euler series, Eq.~\eqref{eq:euler},  with $K=4$ (dashed-dot), $K=20$ (dashed), $K=100$ (solid). Red squares indicate data points from the boundary integral method.}
\label{figeuler}
\end{figure}

The presence of the singularity on the negative real axis for the series $c_k$ led naturally to an Euler transformation to map the singularity to infinity which, as detailed in \S\ref{sec:euler},  yields a series with an infinite radius of convergence in $\delta$, and thus in $\epsilon$. In Fig.~\ref{figeuler} we plot the results of the Euler-transformed series, Eq.~\eqref{eq:euler}, 
 for the swimming speed, $U$, against the wave amplitude, $\epsilon$.  The results are markedly improved over the original unaltered series. With $K=4$ we obtain results which are accurate for up to $\epsilon\approx1.3$, already higher than for Taylor's fourth order formula. With $K=20$ terms, $U(\epsilon)$ is found to be accurate up to $\epsilon\approx2$, and when using $K=100$ terms we obtain results which are accurate for $\epsilon>7$. With all $K=500$ terms the series is accurate up to $\epsilon\approx 15$ with a relative error of 1\% (the series is however convergent for all values of $\epsilon$).

\subsection{Pad\'{e} approximants and Shanks transformation}
\begin{figure}
\centering
\includegraphics[width=0.5\textwidth]{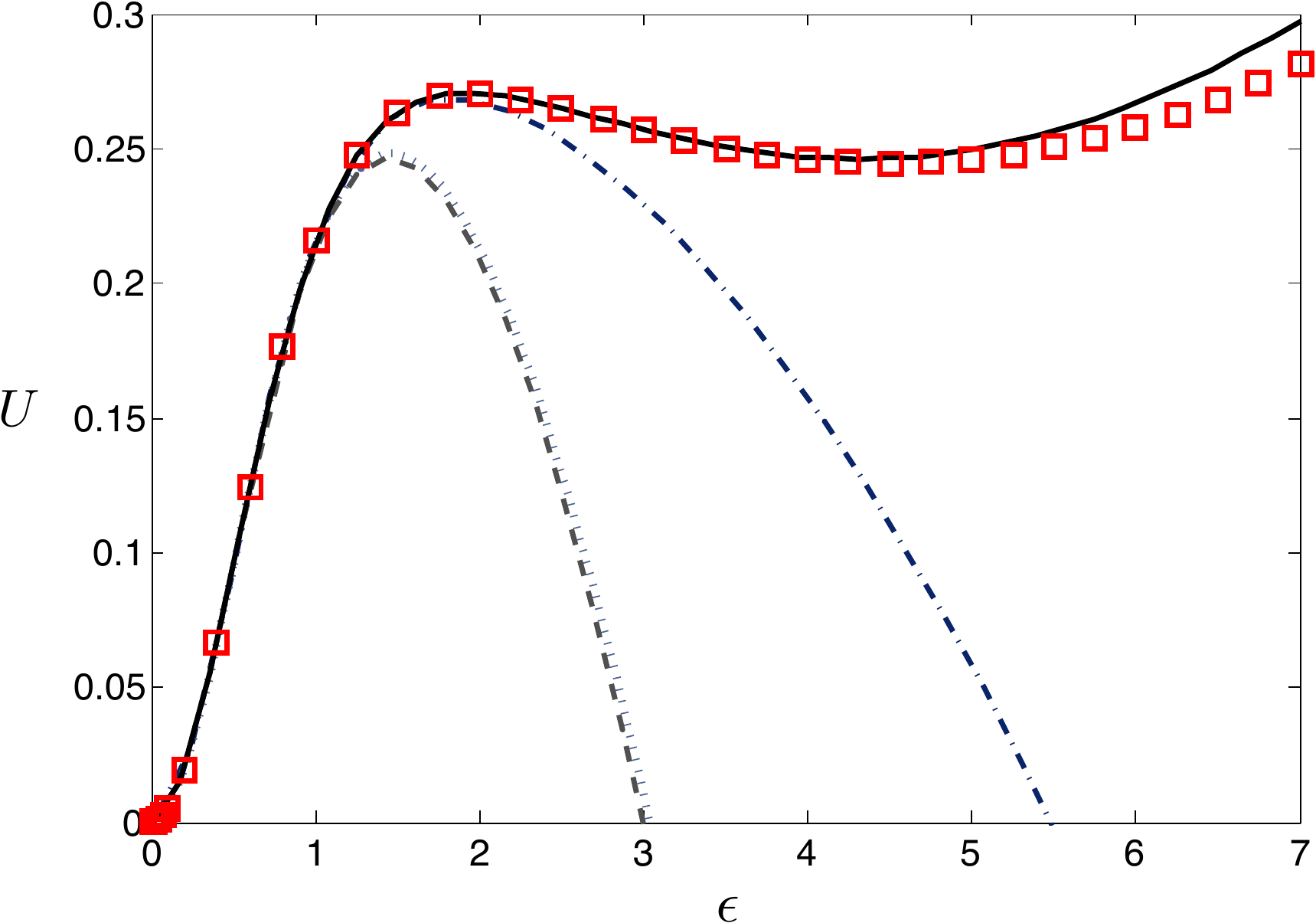}
\caption{Swimming speed, $U$, against amplitude, $\epsilon$, for (repeated) Shanks transformations of partial sums up to $S_2$ (dotted) and $S_6$ (dashed-dot) and for the Pad\'{e} approximants $P_{2}^{2}$ (dashed), $P_{10}^{10}$ (solid). Red squares indicate data points from the boundary integral method.}
\label{figpadeshanks}
\end{figure}

Pad\'{e} approximants provide a convenient (yet brute-force) way to drastically improve the performance of the series without the need to investigate the analytic structure of the underlying function. We find that using only a few terms provides very good results,  as  we show in Fig.~\ref{figpadeshanks}. For $K=4$ we obtain $P_{2}^{2}$ (dashed) which is accurate past the singularity, while for $K=22$ we obtain $P_{10}^{10}$ (solid) which is accurate up to $\epsilon\approx 4$, and shows an error which is reasonably small for larger amplitudes. Unfortunately the coefficient matrix which must be inverted to obtain the $b_k$ coefficients of the Pad\'{e} approximants becomes increasingly ill-conditioned as more terms of the series are added and we see diminishing returns from the Pad\'{e} approximants of higher order expansions; for example, $P_{150}^{150}$ is only accurate up to $\epsilon\approx5$.

Similarly, repeated Shanks transformations of the first few partial sums results in a marked improvement of the convergence of the series. We see in Fig.~\ref{figpadeshanks} that the (repeated) Shanks transformation of partial sums up to $S_2$ (dotted line) yields results nearly identical to the $P_{2}^{2}$ approximant, while for terms up to $S_6$ (dashed-dot) we see reasonable accuracy up to $\epsilon\approx2$ in agreement with the results from Ref.~\cite{drummond66}. We find however that the addition of any further terms in the sequence leads to a pronounced decrease in the convergence properties of the sum.

\section{Concluding Remarks}
Despite its simplicity, Taylor's swimming sheet model is still used to provide physical insight into many interesting natural phenomena. In this paper, we demonstrated that by systematizing the perturbation expansion outlined by Taylor in the wave amplitude, $\epsilon$, the solution for the swimming speed can be obtained in a straightforward fashion to arbitrarily high order. The series unfortunately diverges for $\epsilon\approx 0.9565$ due to a nonphysical first-order pole located  in the negative real axis.  In order to increase the convergence of the series, the singularity can be mapped to infinity via an Euler transformation. The recast series then has an infinite radius of convergence and produces spectacularly  accurate results for very large amplitudes (albeit requiring a good number of terms). An alternative is to reformulate the series using Pad\'{e} approximants or repeated Shanks transformations,  which give reasonable accuracy for moderate amplitudes with fewer terms, but can become problematic for very large amplitudes.

\vspace{1cm}
This paper is dedicated to Steve Childress whose textbook on swimming and flying remains an inspiration.  We thank Glenn Ierley for useful discussions and advice. Funding by the NSF (CBET-0746285) and NSERC (PGS D3-374202) is gratefully acknowledged.

\bibliographystyle{elsarticle-num}

\end{document}